\documentclass[letterpaper,twocolumn,showpacs,
preprintnumbers,amsmath,amssymb,floatfix,nofootinbib]{revtex4}

\usepackage{graphicx}
\usepackage{bm}
\usepackage{url,multirow,xcolor}

\begin{document}

\title{ Overlap of quasiparticle random-phase approximation states for nuclear matrix elements of the neutrino-less double beta decay }

\author{J.\ Terasaki}
\affiliation{Center for Computational Sciences, University of Tsukuba, Tsukuba 305-8577, Japan}

\begin{abstract} 
Quasiparticle random-phase approximation (QRPA) is applied to two nuclei, and overlap of the QRPA excited states based on the different nuclei 
is calculated.
The aim is to calculate the overlap of intermediate nuclear states of the double-beta decay. 
We use the like-particle QRPA after the closure approximation is applied to the nuclear matrix elements. 
The overlap is calculated rigorously by making use of the explicit equation of the QRPA ground state. 
The formulation of the overlap is shown, and a test calculation is performed. 
The effectiveness of the truncations used is shown. 
\end{abstract}

%
\maketitle
%
Needless to say, neutrino physics is important for particle physics 
as it provides information on lepton-number violation, Majorana nature of neutrino, and neutrino mass, see e.g.~\cite{Bil87,Doi85}, 
which are the aspects beyond the Standard Model. 
Neutrino physics is also very interesting for nuclear physics \cite{Fae12,Ver02,Fae98,Suh98,Sim98,Pan94,Tom91,Hax84}, because 
an accurate calculation of nuclear matrix elements of the neutrino-less double-beta (0$\nu\beta\beta$) decay is necessary, e.g.~\cite{Doi85,Hax84}, 
along with the experimental half-life value of the initial nucleus in order to determine the neutrino mass in one of few methods. 
This calculation is a challenging opportunity to test the capability of the techniques of theoretical nuclear physics for calculating many-body correlations in heavy nuclei.

One of the important theoretical methods to calculate the nuclear matrix elements is the proton-neutron quasiparticle random-phase approximation (pn-QRPA), 
a method that has been improved significantly in the past few decades. 
A few of the milestones include the correct treatment of the effects of the Pauli-exclusion principle 
in the intermediate states by the renormalized pn-QRPA \cite{Cat94}, fulfilment of the Ikeda sum rule by the fully renormalized pn-QRPA \cite{Pac03}, 
inclusion of the pn-pairing correlations \cite{Sch96}, and extension to the deformed states \cite{Sim04}.
Nevertheless, the nuclar matrix elements calculated by various approaches including the pn-QRPA are distributed in a range of a factor of 2 \cite{Fae12,MED11}.

We calculate the nuclear matrix elements of the $0\nu\beta\beta$ decay by employing the like-particle QRPA \cite{Vog10,Rin80}, 
which can be applied after the closure approximation \cite{Tom91,Fae91,Suh91,Pan90,Hax84}  is used. 
The advantage of this approach is that the intermediate states are free from the Pauli-exclusion-principle problem without any modification 
because they are states of even-even nuclei. 
In addition, we calculate the overlap of the intermediate states obtained by the like-particle QRPA with greater accuracy than that in the previous studies 
 \cite{Sim04,Kam91,Sim98,Civ86,Gro85}. 
It is emphasized that the importance of the overlap of the intermediate states has been pointed out in Ref.~\cite{Sim04} in terms of deformation. 
The purpose of this study is to demonstrate the feasibility of the overlap calculation using the ground-state wave function of the like-particle QRPA 
\textit{explicitly}.
The equation of the QRPA ground state has been known for decades, e.g.~\cite{Bal69}. 
To the best of our knowledge, however, this is the first time that a numerical calculation has been carried out rigorously. 

The axial and parity symmetries of the nuclei are assumed throughout this paper.
The $z$-component of the angular momentum is denoted by $j^z_\alpha$ for nucleon state $\alpha$ and by $K_m$ for nuclear state $m$.
$\pi_\alpha$ and $\pi_m$ are used to denote the parity.
Hereafter, we refer to the like-particle QRPA as simply QRPA. 
After the closure approximation is applied to the nuclear matrix elements of the $0\nu\beta\beta$ decay, e.g.~Refs.~\cite{Sim99,Doi85,Hax84},
one of those matrix elements, the Gamow-Teller type as an example, is written as
\begin{eqnarray}
M^{(0\nu)}_\textrm{GT} &=&
\langle F|\sum_{ij}h_+(r_{ij},\bar{E}_a)\bm{\sigma}(i)\cdot\bm{\sigma}(j)\tau_{+}(i)\tau_{+}(j) | I \rangle \nonumber \\
&=& \sum_{\alpha\beta}\sum_{\alpha^\prime\beta^\prime}
\langle \alpha \alpha^\prime| h_+(r_{12},\bar{E}_a)\bm{\sigma}(1)\cdot \bm{\sigma}(2) \nonumber\\
&&\times \tau_{+}(1) \tau_{+}(2) | \beta \beta^\prime \rangle 
\sum_{mm^\prime}\langle F | c^\dagger_{\alpha} c^\dagger_{\alpha^\prime} O^{F\dagger}_m | F \rangle \nonumber \\
&&\times\langle F | O^F_m O^{I\dagger}_{m^\prime} | I \rangle \langle I | O^I_{m^\prime} c_{\beta^\prime} c_{\beta} | I \rangle , \label{eq:double_GT_our_original}
\end{eqnarray}
where $|F\rangle$ and $|I\rangle$ denote the final and initial nuclear states of the decay, and the ground states of the QRPA are used. 
$h_+(r_{ij},\bar{E}_a)$ is the neutrino potential \cite{Doi85} with $r_{ij}=|\bm{r}_i-\bm{r}_j|$, and 
$\bar{E}_a$ being the average energy of the intermediate nuclear states. 
$i$ $(j)$ indicates a nucleon. 
$\bm{\sigma}$ denotes the spin-Pauli matrix, $\tau_+$ is the raising operator of the $z$-component of 
the isospin. An arbitrary single-particle basis $\{\alpha\}$ is introduced, 
and the creation and annihilation operators of the single-particle state are denoted by $c^\dagger_{\alpha}$ and $c_\alpha$, respectively.
The creation and annihilation operators of the excited state $m$ of the QRPA based on the initial state
are denoted by $O^{I\dagger}_m$ and $O^I_m$, respectively,  
and those based on the final state are denoted by $O^{F\dagger}_m$ and $O^F_m$, respectively.
The completeness using these operators is used in Eq.~(\ref{eq:double_GT_our_original}).
There is no selection rule for the intermediate states with respect to $(K_m \pi_m)$.
Simplified notations $K=K_m$ and $\pi=\pi_m$ are used hereafter.

We express $|I\rangle$ in the form \cite{Bal69}
\begin{eqnarray}
&& |I\rangle = \frac{1}{ {\cal N}_I} \prod_{K^\prime \pi^\prime} \exp\left[\hat{v}^{(K^\prime \pi^\prime)}_I\right]|i\rangle,
\end{eqnarray}
where
$|i\rangle$ is the HFB ground state of the same nucleus as the one $|I\rangle$ describes, and 
$\hat{v}^{(K^\prime\pi^\prime)}_I$ is a generator of the QRPA ground state. 
${\cal N}_I$ is the normalization factor.
In this paper, for any equation only referring to the initial state denoted by $I$ or $i$, we have also provided the corresponding equation referring to $F$ or $f$.  Latter equations are omitted. We have $[O^{I\dagger}_m,O^I_{m^\prime}]=0$ in the QRPA, if $(K\pi)\neq (K_{m^\prime}\pi_{m^\prime})$; 
hence, $\hat{v}^{(K^\prime\pi^\prime)}_I$'s with different $(K^\prime\pi^\prime)$ are separately determined by
\begin{eqnarray}
&&O^I_{m^\prime} \exp \left[ \hat{v}^{(K_{m^\prime}\pi_{m^\prime})}_I \right] |i \rangle = 0.
\end{eqnarray}

A general quasiparticle basis $\{\mu\}$ based on the initial state is introduced by 
$ a^I_\mu|i\rangle = 0 $, 
where the $a^I_\mu$ is annihilation operator, $\mu=(q_\mu,\pi_\mu,j^z_\mu, i_\mu)$ being the label of a general quasiparticle state. 
$q_\mu$ denotes a proton or neutron, and $i_\mu$ is the label specifying a state in the subspace ($q_\mu,\pi_\mu,j^z_\mu$). 
Notation $-\mu$ is used for expressing $(q_\mu,\pi_\mu,-j^z_\mu,i_\mu)$. 
We use the canonical-quasiparticle basis \cite{Rin80} $\{\mu\}$ for efficiency of the QRPA calculation \cite{Ter10}.
$\hat{v}^{(K^\prime\pi^\prime)}_I$ is expressed as  
\begin{eqnarray}
&&\hat{v}^{(K^\prime\pi^\prime)}_I = \sum_{\mu\nu\mu^\prime\nu^\prime} C^{(K^\prime\pi^\prime)I}_{\mu\nu,\mu^\prime\nu^\prime}
 a^{I\dagger}_\mu a^{I\dagger}_\nu a^{I\dagger}_{\mu^\prime} a^{I\dagger}_{\nu^\prime}. \label{eq:vhatI}
\end{eqnarray}
$a^{I\dagger}_\mu a^{I\dagger}_\nu$ and $a^{I\dagger}_{\mu^\prime} a^{I\dagger}_{\nu^\prime}$ 
in Eq.~(\ref{eq:vhatI}) are the fermion image of the boson \cite{Bal69};  
a condition is introduced that $C^{(K^\prime\pi^\prime)I}_{\mu\nu,\mu^\prime\nu^\prime}$ does not vanish, only if 
$j^z_\mu+j^z_\nu=K^\prime$, $j^z_{\mu^\prime}+j^z_{\nu^\prime}=-K^\prime$, and 
$\pi_\mu \pi_\nu= \pi_{\mu^\prime}\pi_{\nu^\prime}=\pi^\prime$. 
We order the canonical-quasiparticle states and restrict $\mu<\nu$, $\mu^\prime<\nu^\prime$ in $C^{(K^\prime\pi^\prime)I}_{\mu\nu,\mu^\prime\nu^\prime}$  
without loss of generality. 

The solution of the QRPA equation gives us 
\begin{eqnarray}
&& O^{I\dagger}_{m^\prime} = \sum_{\mu < \nu}\left( X^{Im^\prime}_{\mu\nu} a^{I\dagger}_\mu a^{I\dagger}_\nu 
 - Y^{Im^\prime}_{-\mu-\nu} a^I_{-\nu} a^I_{-\mu} \right)~,
\label{eq:O_QRPA}
\end{eqnarray}
where $j^z_\mu+j^z_\nu = K_{m^\prime}$ and $\pi_\mu \pi_\nu=\pi_{m^\prime}$.
We define matrices
\begin{eqnarray}
 &&C^{(K^\prime\pi^\prime)I} = \left( 
 \begin{array}{ccc}
  C^{(K^\prime\pi^\prime)I}_{11,-1-1} & \cdots & C^{(K^\prime\pi^\prime)I}_{11,-n-n^\prime} \\
   & \cdots & \\
  C^{(K^\prime\pi^\prime)I}_{nn^\prime,-1-1} & \cdots & C^{(K^\prime\pi^\prime)I}_{nn^\prime,-n-n^\prime}
 \end{array}
 \right)~,
\\
\nonumber\\
 &&X^{(K^\prime\pi^\prime)I} = \left( 
 \begin{array}{ccc}
  X^{I1}_{11} & \cdots & X^{IM}_{11} \\
    & \cdots & \\
  X^{I1}_{nn^\prime} & \cdots & X^{IM}_{nn^\prime}
 \end{array}
 \right),
\end{eqnarray}
%
where the QRPA solutions having $(K^\prime\pi^\prime)$ are used. 
The negative integers of the index correspond to $-\mu$.
Matrices $Y^{(K^\prime\pi^\prime)I}$, 
$C^{(K^\prime\pi^\prime)F}$, $X^{(K^\prime\pi^\prime)F}$, and $Y^{(K^\prime\pi^\prime)F}$ are also introduced in the same way.
$C^{(K^\prime\pi^\prime)I}$ is obtained, by ignoring the exchange terms (the quasi-boson approximation), as follows:
\begin{eqnarray}
 C^{(K^\prime\pi^\prime)I} = \frac{1}{1+\delta_{K0}}
\left( Y^{(K^\prime\pi^\prime)I} \frac{1}{X^{(K^\prime\pi^\prime)I}} \right)^\textrm{T},
\end{eqnarray}
where suffix T stands for transpose. Practically, $1/X^{(K^\prime\pi^\prime)I}$ does not have a singularity.

The relation between the two HFB states can be written as \cite{Rin80} 
\begin{eqnarray}
|i\rangle = \frac{1}{{\cal N}_i} \exp\left[ \sum_{\mu\nu} D_{\mu\nu} a^{F\dagger}_\mu a^{F\dagger}_\nu \right] | f\rangle~, \label{eq:i-f} \\
{\cal N}_i = \frac{1}{\langle f|i\rangle} = \sqrt{\det(I+D^\dagger D)}~. \label{eq:N_HFB}
\end{eqnarray}
$I$ is the unit matrix of the size of matrix $D$, which is defined by
\begin{eqnarray}
\left( D \right)_{ij}=D_{\mu_i -\mu_j},\ i,j=1,\cdots,n_T,\label{eq:D}
\end{eqnarray}
$n_T$ is the dimension of the subspace $(q_\mu,\pi_\mu,j^z_\mu)$.
$D_{\mu\nu}$ is not equal to 0 only for those $\mu$ and $\nu$ that satisfy $j^z_\mu+j^z_\nu = 0 $ and 
$\pi_\mu \pi_\nu = +$. We restrict $j^z_\mu>0$ in Eq.~(\ref{eq:i-f}).
\noindent
The unitary transformation from basis $\{a^{F\dagger}_\mu, a^{F}_{-\mu} \}$ to basis $\{ a^{I\dagger}_\mu, a^{I}_{-\mu} \}$ is given by
\begin{eqnarray}
a^{I\dagger}_\mu = \sum_{\mu^\prime} \left(
 T^{IF1}_{\mu\mu^\prime} a^{F\dagger}_{\mu^\prime} + T^{IF2}_{\mu-\mu^\prime} a^F_{-\mu^\prime} \right)~, \label{eq:transformation_qp}
\end{eqnarray}
and its Hermite conjugate equation for $-\mu$ with $j^z_\mu=j^z_{\mu^\prime}$ and $\pi_\mu=\pi_{\mu^\prime}$. 
$T^{IF1}_{\mu\mu^\prime}$ and $T^{IF2}_{\mu-\mu^\prime}$ can be obtained from the volume integral of the product of 
the canonical-quasiparticle wave functions \cite{Dob84} of the two bases.  
$D_{\mu -\nu}$ is given by
\begin{eqnarray}
&&D = -\left( \frac{1}{T^{IF1}}T^{IF2}\right)^\ast~, \label{eq:D} \\
&&\left( T^{IF1}\right)_{ij}=T^{IF1}_{\mu_i \mu_j}, \ i,j=1,\cdots,n_T. 
\end{eqnarray}
Matrix $T^{IF2}$ is defined in the same way as matrix $D$.
Practically, again, $1/T^{IF1}$ does not have a singularity. 

Now, we expand and truncate the overlap matrix element with respect to $\hat{v}^{(K^\prime\pi^\prime)}_F$ and 
$\hat{v}^{(K^\prime\pi^\prime)}_I$ as
\begin{eqnarray}
_F\langle m | m^\prime \rangle_I &\equiv& 
\langle F | O^F_m O^{I\dagger}_{m^\prime} |I\rangle \nonumber \\
&\simeq& 
\frac{1}{ {\cal N}_I {\cal N}_F } 
\left( G^{FI0}_{mm^\prime} +G^{FI1}_{mm^\prime} +G^{FI2}_{mm^\prime} \right), 
 \label{eq:fooi}
\end{eqnarray}
\begin{eqnarray}
G^{FI0}_{mm^\prime}&=&\langle f | O^F_m O^{I\dagger}_{m^\prime} | i \rangle, \label{eq:GFI0} \\
G^{FI1}_{mm^\prime}&=&\sum_{K_1 \pi_1} \Big( \langle f | \hat{v}^{(K_1\pi_1)\dagger}_F O^F_m O^{I\dagger}_{m^\prime} | i \rangle 
\nonumber \\
&&+ \langle f | O^F_m O^{I\dagger}_{m^\prime} \hat{v}^{(K_1\pi_1)}_I | i \rangle \Big), \label{eq:GFI1} \\
G^{FI2}_{mm^\prime}&=&\sum_{K_1\pi_1} \langle f | \hat{v}^{(K_1\pi_1)\dagger}_F O^F_m O^{I\dagger}_{m^\prime} \hat{v}^{(K_1\pi_1)}_I | i \rangle, 
\label{eq:GFI2}
\end{eqnarray}
%
\begin{eqnarray}
{\cal N}_I &\simeq& \bigg[ 1+\sum_{K_1\pi_1} \bigg\{ \langle i | \hat{v}^{(K_1\pi_1)\dagger}_I 
\hat{v}^{(K_1\pi_1)}_I | i \rangle \nonumber \\  
&&+\frac{1}{4} \langle i | \left( \hat{v}^{(K_1\pi_1)\dagger}_I \right)^2 \left( \hat{v}^{(K_1\pi_1)}_I \right)^2 | i \rangle 
\bigg\} \bigg]^{1/2}. \label{eq:NI_QRPA}
\end{eqnarray}
We test up to the second-order terms $G^{FI2}_{mm^\prime}$, 
which use both $\hat{v}_F^{(K_1\pi_1)}$ and $\hat{v}_I^{(K_2\pi_2)}$, but only with $(K_1\pi_1)= (K_2\pi_2)$ in Eq.~(\ref{eq:fooi}) 
(actually $G^{FI2}_{mm^\prime}$ is negligible in most of the overlap matrix elements, as shown later). 
Up to the fourth-order terms are included in normalization factors ${\cal N}_I$ and ${\cal N}_F$, 
because its convergence of the $\hat{v}$-expansion is slow as compared to the un-normalized overlap matrix elements 
(which are the result of the numerical test).  
Equations (\ref{eq:GFI0})$-$(\ref{eq:GFI2}) can be calculated using $X^{Im^\prime}_{\mu\nu}$, $Y^{Im^\prime}_{-\mu-\nu}$, 
$C^{(K^\prime\pi^\prime)I}_{\mu\nu,\mu^\prime\nu^\prime}$, those of $F$, $T^{IF1}_{\mu\mu^\prime}$, $T^{IF2}_{\mu-\mu^\prime}$ and 
$D_{\mu\nu}$. The concrete equations will be given in the forthcoming full paper.

We use the code of the HFB approximation \cite{Bla05} and that of the QRPA developed by us \cite{Ter10}. 
The wave functions are treated, in both the codes, on a mesh within the cylindrical box 
and are discretized by the vanishing boundary condition at the edge of the box. 
The HFB equation is solved using a cutoff of the quasiparticle energy at 20 MeV for convenience in performing the tests. 
We transform the wave functions of the quasiparticle states to those of the canonical-quasiparticle states before the QRPA calculation \cite{Ter05}.

$^{26}$Mg and $^{26}$Si are used in this paper for $|i\rangle$ ($|I\rangle$) and $|f\rangle$ ($|F\rangle$), respectively, 
with the Skyrme parameter set SkM$^\ast$ \cite{Bar82} and the volume pairing density functional \cite{Dob96}. 
The properties of the HFB ground states are shown in Table \ref{tab:26Mg_26Si}. 
The total dimension of the HFB space is $\simeq$330, including those with negative $j_z$.
Strength of the volume pairing $G_\textrm{n}=-270$ MeV$\:$fm$^3$ is used for the neutrons of both $^{26}$Mg and $^{26}$Si, and  
$G_\textrm{p}=-150.0$  and $-270.0$ MeV$\:$fm$^3$ is used for the protons of $^{26}$Mg and $^{26}$Si, respectively. 
These values were chosen so that the mean fields of the ground states are similar between the two nuclei.
\begin{table}[h]
\caption{\label{tab:26Mg_26Si}%
Properties of the HFB ground states of $^{26}$Mg and $^{26}$Si used in this paper. 
$\beta_\textrm{p}$ and $\Delta_\textrm{p}$ denote, respectively, the quadrupole deformation and the averaged pairing gap of the 
protons. $\beta_\textrm{n}$ and $\Delta_\textrm{n}$ denote the same for the neutrons.}
\begin{ruledtabular}
\begin{tabular}{ccccc}
 nucleus & $\beta_\textrm{p}$ & $\Delta_\textrm{p}$ (MeV) & $\beta_\textrm{n}$ & $\Delta_\textrm{n}$ (MeV) \\
\colrule
$^{26}$Mg & $-0.199$ &0.794 &$-0.195$ & 1.510 \\
$^{26}$Si  & $-0.224$ & 0.865 & $-0.206$ & 1.402 \\
\end{tabular}
\end{ruledtabular}
\end{table}

\begin{figure}[h]
\includegraphics[width=8cm]{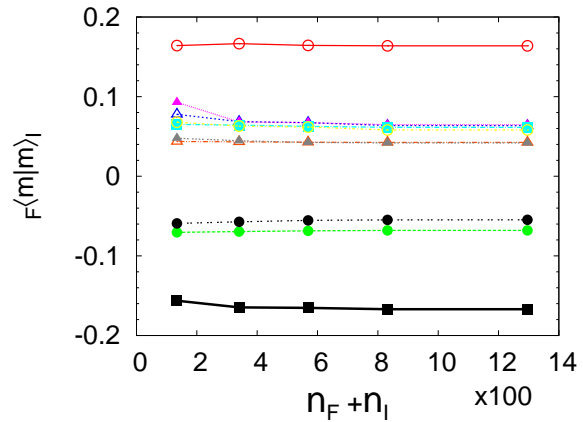} 
\caption{ \label{fig:fNF+fNI} (Color online) Ten diagonal overlap matrix elements having the largest absolute values, as functions of $\mathfrak{N}_F+\mathfrak{N}_I$. }
\end{figure}



\begin{figure}[h]
\includegraphics[width=8cm]{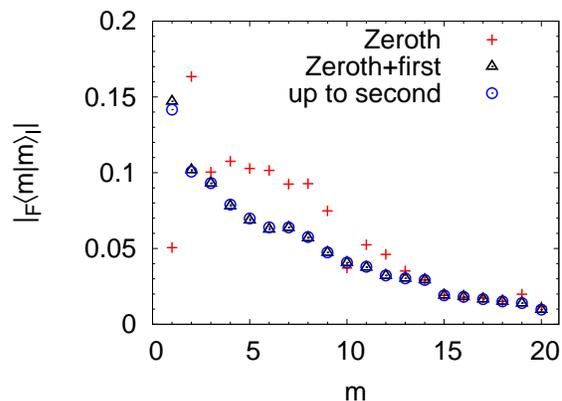}
\caption{ \label{fig:abs_ov_diag_I} (Color online) Twenty largest absolute values of the diagonal overlap matrix elements. Those up to the second order with respect to $\hat{v}^{(0+)}_I$ and $\hat{v}^{(0+)}_F$ are shown in the descending order. 
$(K_1\pi_1)\neq (0+)$ are not included in $G^{FI1}_{mm^\prime}$ (\ref{eq:GFI1}) and $G^{FI2}_{mm^\prime}$ (\ref{eq:GFI2}). 
$\mathfrak{N}_F+\mathfrak{N}_I=134$ was used; see Fig.~\ref{fig:fNF+fNI}.
}
\end{figure}

%

\begin{figure}[h]
\includegraphics[width=8cm]{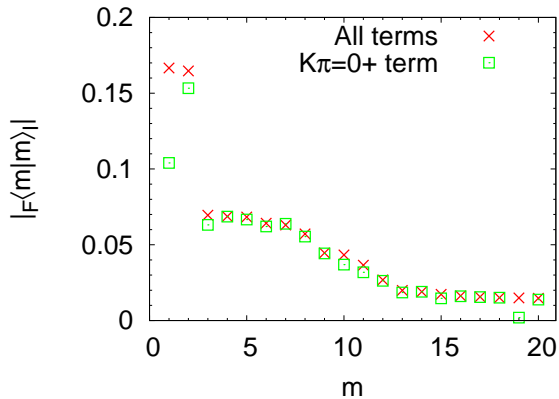}
\caption{ \label{fig:abs_ov_diag_kpnokp} (Color online)  Twenty largest absolute values of the diagonal matrix elements of the overlap. 
$G^{FI1}_{mm^\prime}$ (\ref{eq:GFI1}) was calculated with $(K_1\pi_1) = (0+)-(4+)$ (all terms), and the terms with only (0+) ($K\pi=0+$ term). 
$\mathfrak{N}_F+\mathfrak{N}_I=350$ is used. $G^{FI2}_{mm^\prime}$ is not included.
QRPA solution number $m$ does not necessarily correspond to that of Fig.~\ref{fig:abs_ov_diag_I}. }
\end{figure}



We show the results of the calculation of $(K\pi)=(0+)$ below. 
Let $\mathfrak{N}_F$ and $\mathfrak{N}_I$ be the number of the two-canonical-quasiparticle states associated with $|F\rangle$ and $|I\rangle$, truncated 
by the cutoff occupation probability 
for calculating Eqs.~(\ref{eq:GFI1}) and (\ref{eq:GFI2}) (those with larger occupation probabilities than the cutoff are used). 
This is another truncation after the 20-MeV cutoff.
The convergence of the overlap matrix elements is shown with respect to $\mathfrak{N}_F+\mathfrak{N}_I$ in Fig.~\ref{fig:fNF+fNI}. 
The same value of the cutoff is applied for the two bases, and we have  $\mathfrak{N}_F\simeq\mathfrak{N}_I$. It is seen that $\mathfrak{N}_F+\mathfrak{N}_I=350$ is sufficient for the convergence. 
The total number without the truncation is $\simeq$3300; thus, this truncation is rather efficient.
$|I\rangle$ and $|F\rangle$  have different configurations at the Fermi surface; therefore, the high-energy components of $O^{I\dagger}_{m^\prime}$ and $O^{F\dagger}_m$ leaving the configuration around the Fermi surface intact do not contribute to the overlap matrix elements.
On the other hand, it is necessary to calculate ${\cal N}_I$ (\ref{eq:NI_QRPA}) and ${\cal N}_F$ without this truncation.

The major diagonal overlap matrix elements are shown in Fig.~\ref{fig:abs_ov_diag_I}, obtained with $\mathfrak{N}_F+\mathfrak{N}_I=134$. 
It is observed that the contribution of $G^{FI2}_{mm^\prime}$ (\ref{eq:GFI2}) is negligible 
and that of $G^{FI1}_{mm^\prime}$ (\ref{eq:GFI1}) is not  significant for the small matrix elements. 
$G^{FI0}_{mm^\prime}$ (\ref{eq:GFI0}) is sufficient in most of the matrix elements omitted in that figure.

The contribution of $(K_1\pi_1) \neq (0+)$ to the major overlap matrix elements through $G^{FI1}$ (\ref{eq:GFI1}) is shown in Fig.~\ref{fig:abs_ov_diag_kpnokp}, calculated with $\mathfrak{N}_F+\mathfrak{N}_I=350$ and max $|K|$ = 4. 
We also calculated the contribution of $(K_1\pi_1)=(0-)$ and $(1-)$ and found that it was smaller than that of the positive parity by at least an order of magnitude; 
thus, only the positive parity is  used. 
The contribution of $(K_1\pi_1) \neq (0+)$ is very small to all of $G^{FI1}$ except for $m=1$, which is one of the spurious states. 
Actually, our method should be applied only to the cases that do not have large fluctuations of the particle number so that the spurious states are not crucial to  the nuclear matrix elements (\ref{eq:double_GT_our_original}).
${\cal N}_F$ and ${\cal N}_I$ require $|K|$ of up to 3 with both parities.

In summary, the overlap matrix elements of the QRPA states based on the ground states of different nuclei have been calculated using the QRPA ground states explicitly for relatively light nuclei with the Skyrme and the contact volume pairing energy functionals. 
The most important finding of this study is that the bold truncations are allowed in the calculation of the un-normalized overlap matrix elements. 
The normalization factors need a less-truncated calculation; however, 
the amount of this calculation is reduced tremendously by identifying $|f\rangle$ and $|i\rangle$ in each factor. 
Considering this advantage and the performance of the modern parallel computers, we believe that there is no reason to avoid the explicit wave function of
the QRPA ground states.

We are grateful to Dr.~Engel for suggesting the application of the like-particle QRPA to the nuclear matrix elements of the $0\nu\beta\beta$ decay.
We thank Drs. Oberacker and Umar for letting us use their HFB code.
This research is supported by the Ministry of Education, Culture, Sports, Science and Technology-Japan under the Grant-in-Aid for Scientific Research No.~23840005. 
Computers were used at the Center for Computational Sciences, University of Tsukuba, under the Collaborative Interdisciplinary Program and the Computational Fundamental Science Project (T2K-Tsukuba); Yukawa Institute for Theoretical Physics, Kyoto University (SR16000); and Research Center for Nuclear Physics, Osaka University 
(SX-8R) . 
Computers were also used at the National Institute for Computational
Sciences and the National Energy Research Scientific Computing Center at the early stage of this research.

\begin{thebibliography}{31}
\expandafter\ifx\csname natexlab\endcsname\relax\def\natexlab#1{#1}\fi
\expandafter\ifx\csname bibnamefont\endcsname\relax
  \def\bibnamefont#1{#1}\fi
\expandafter\ifx\csname bibfnamefont\endcsname\relax
  \def\bibfnamefont#1{#1}\fi
\expandafter\ifx\csname citenamefont\endcsname\relax
  \def\citenamefont#1{#1}\fi
\expandafter\ifx\csname url\endcsname\relax
  \def\url#1{\texttt{#1}}\fi
\expandafter\ifx\csname urlprefix\endcsname\relax\def\urlprefix{URL }\fi
\providecommand{\bibinfo}[2]{#2}
\providecommand{\eprint}[2][]{\url{#2}}

\bibitem[{\citenamefont{Bilenky and Petcov}(1987)}]{Bil87}
\bibinfo{author}{\bibfnamefont{S.~M.} \bibnamefont{Bilenky}} \bibnamefont{and}
  \bibinfo{author}{\bibfnamefont{S.~T.} \bibnamefont{Petcov}},
  \bibinfo{journal}{Rev. Mod. Phys.} \textbf{\bibinfo{volume}{59}},
  \bibinfo{pages}{671} (\bibinfo{year}{1987}).

\bibitem[{\citenamefont{Doi et~al.}(1985)\citenamefont{Doi, Kotani, and
  Takasugi}}]{Doi85}
\bibinfo{author}{\bibfnamefont{M.}~\bibnamefont{Doi}},
  \bibinfo{author}{\bibfnamefont{T.}~\bibnamefont{Kotani}}, \bibnamefont{and}
  \bibinfo{author}{\bibfnamefont{E.}~\bibnamefont{Takasugi}},
  \bibinfo{journal}{Prog. Theor. Phys. Suppl.} \textbf{\bibinfo{volume}{83}},
  \bibinfo{pages}{1} (\bibinfo{year}{1985}).

\bibitem[{\citenamefont{Faessler}(2012)}]{Fae12}
\bibinfo{author}{\bibfnamefont{A.}~\bibnamefont{Faessler}},
  \bibinfo{howpublished}{e-print arXiv:nucl-th/1203.3648}
  (\bibinfo{year}{2012}).

\bibitem[{\citenamefont{Vergados}(2002)}]{Ver02}
\bibinfo{author}{\bibfnamefont{J.~D.} \bibnamefont{Vergados}},
  \bibinfo{journal}{Phys. Rep.} \textbf{\bibinfo{volume}{361}},
  \bibinfo{pages}{1} (\bibinfo{year}{2002}).

\bibitem[{\citenamefont{Faessler and \v{S}imkovic}(1998)}]{Fae98}
\bibinfo{author}{\bibfnamefont{A.}~\bibnamefont{Faessler}} \bibnamefont{and}
  \bibinfo{author}{\bibfnamefont{F.}~\bibnamefont{\v{S}imkovic}},
  \bibinfo{journal}{J. Phys. G} \textbf{\bibinfo{volume}{24}},
  \bibinfo{pages}{2139} (\bibinfo{year}{1998}).

\bibitem[{\citenamefont{Suhonen and Civitarese}(1998)}]{Suh98}
\bibinfo{author}{\bibfnamefont{J.}~\bibnamefont{Suhonen}} \bibnamefont{and}
  \bibinfo{author}{\bibfnamefont{O.}~\bibnamefont{Civitarese}},
  \bibinfo{journal}{Phys. Rep.} \textbf{\bibinfo{volume}{300}},
  \bibinfo{pages}{123} (\bibinfo{year}{1998}).

\bibitem[{\citenamefont{\v{S}imkovic et~al.}(1998)\citenamefont{\v{S}imkovic,
  Pantis, and Faessler}}]{Sim98}
\bibinfo{author}{\bibfnamefont{F.}~\bibnamefont{\v{S}imkovic}},
  \bibinfo{author}{\bibfnamefont{G.}~\bibnamefont{Pantis}}, \bibnamefont{and}
  \bibinfo{author}{\bibfnamefont{A.}~\bibnamefont{Faessler}},
  \bibinfo{journal}{Prog. Part. Nucl. Phys.} \textbf{\bibinfo{volume}{40}},
  \bibinfo{pages}{285} (\bibinfo{year}{1998}).

\bibitem[{\citenamefont{Pantis and Vergados}(1994)}]{Pan94}
\bibinfo{author}{\bibfnamefont{G.}~\bibnamefont{Pantis}} \bibnamefont{and}
  \bibinfo{author}{\bibfnamefont{J.~D.} \bibnamefont{Vergados}},
  \bibinfo{journal}{Phys. Rep.} \textbf{\bibinfo{volume}{242}},
  \bibinfo{pages}{285} (\bibinfo{year}{1994}).

\bibitem[{\citenamefont{Tomoda}(1991)}]{Tom91}
\bibinfo{author}{\bibfnamefont{T.}~\bibnamefont{Tomoda}},
  \bibinfo{journal}{Rep. Prog. Phys.} \textbf{\bibinfo{volume}{54}},
  \bibinfo{pages}{53} (\bibinfo{year}{1991}).

\bibitem[{\citenamefont{Haxton and \hbox{Stephenson Jr.}}(1984)}]{Hax84}
\bibinfo{author}{\bibfnamefont{W.~C.} \bibnamefont{Haxton}} \bibnamefont{and}
  \bibinfo{author}{\bibfnamefont{G.~J.} \bibnamefont{\hbox{Stephenson Jr.}}},
  \bibinfo{journal}{Prog. Part. Nucl. Phys.} \textbf{\bibinfo{volume}{12}},
  \bibinfo{pages}{409} (\bibinfo{year}{1984}).

\bibitem[{\citenamefont{Catara et~al.}(1994)\citenamefont{Catara, Dang, and
  Sambataro}}]{Cat94}
\bibinfo{author}{\bibfnamefont{F.}~\bibnamefont{Catara}},
  \bibinfo{author}{\bibfnamefont{N.~D.} \bibnamefont{Dang}}, \bibnamefont{and}
  \bibinfo{author}{\bibfnamefont{M.}~\bibnamefont{Sambataro}},
  \bibinfo{journal}{Nucl. Phys. A} \textbf{\bibinfo{volume}{579}},
  \bibinfo{pages}{1} (\bibinfo{year}{1994}).

\bibitem[{\citenamefont{Pacearescu et~al.}(2003)\citenamefont{Pacearescu,
  Rodin, \v{S}imkovic, and Faessler}}]{Pac03}
\bibinfo{author}{\bibfnamefont{L.}~\bibnamefont{Pacearescu}},
  \bibinfo{author}{\bibfnamefont{V.}~\bibnamefont{Rodin}},
  \bibinfo{author}{\bibfnamefont{F.}~\bibnamefont{\v{S}imkovic}},
  \bibnamefont{and} \bibinfo{author}{\bibfnamefont{A.}~\bibnamefont{Faessler}},
  \bibinfo{journal}{Phys. Rev. C} \textbf{\bibinfo{volume}{68}},
  \bibinfo{pages}{064310} (\bibinfo{year}{2003}).

\bibitem[{\citenamefont{Schwieger et~al.}(1996)\citenamefont{Schwieger,
  \v{S}imkovic, and Faessler}}]{Sch96}
\bibinfo{author}{\bibfnamefont{J.}~\bibnamefont{Schwieger}},
  \bibinfo{author}{\bibfnamefont{F.}~\bibnamefont{\v{S}imkovic}},
  \bibnamefont{and} \bibinfo{author}{\bibfnamefont{A.}~\bibnamefont{Faessler}},
  \bibinfo{journal}{Nucl. Phys. A} \textbf{\bibinfo{volume}{600}},
  \bibinfo{pages}{179} (\bibinfo{year}{1996}).

\bibitem[{\citenamefont{\v{S}imkovic et~al.}(2004)\citenamefont{\v{S}imkovic,
  Pacearescu, and Faessler}}]{Sim04}
\bibinfo{author}{\bibfnamefont{F.}~\bibnamefont{\v{S}imkovic}},
  \bibinfo{author}{\bibfnamefont{L.}~\bibnamefont{Pacearescu}},
  \bibnamefont{and} \bibinfo{author}{\bibfnamefont{A.}~\bibnamefont{Faessler}},
  \bibinfo{journal}{Nucl. Phys. A} \textbf{\bibinfo{volume}{733}},
  \bibinfo{pages}{321} (\bibinfo{year}{2004}).

\bibitem[{MED()}]{MED11}
\bibinfo{howpublished}{[http://medex11.utef.cvut.cz]}.

\bibitem[{\citenamefont{Vogel}(2010)}]{Vog10}
\bibinfo{author}{\bibfnamefont{P.}~\bibnamefont{Vogel}},
  \emph{\bibinfo{title}{Current Aspects of Neutrino Physics, edited by D. O.
  Caldwell}} (\bibinfo{publisher}{Springer-Verlag}, \bibinfo{address}{New
  York}, \bibinfo{year}{2010}), p. \bibinfo{pages}{177}.

\bibitem[{\citenamefont{Ring and Schuck}(1980)}]{Rin80}
\bibinfo{author}{\bibfnamefont{P.}~\bibnamefont{Ring}} \bibnamefont{and}
  \bibinfo{author}{\bibfnamefont{P.}~\bibnamefont{Schuck}},
  \emph{\bibinfo{title}{The Nuclear Many-body Problem}}
  (\bibinfo{publisher}{Springer-Verlag}, \bibinfo{address}{Berlin},
  \bibinfo{year}{1980}).

\bibitem[{\citenamefont{Faessler et~al.}(1991)\citenamefont{Faessler,
  Kami\'{n}ski, Pantis, and Vergados}}]{Fae91}
\bibinfo{author}{\bibfnamefont{A.}~\bibnamefont{Faessler}},
  \bibinfo{author}{\bibfnamefont{W.~A.} \bibnamefont{Kami\'{n}ski}},
  \bibinfo{author}{\bibfnamefont{G.}~\bibnamefont{Pantis}}, \bibnamefont{and}
  \bibinfo{author}{\bibfnamefont{J.~D.} \bibnamefont{Vergados}},
  \bibinfo{journal}{Phys. Rev. C} \textbf{\bibinfo{volume}{43}},
  \bibinfo{pages}{R21} (\bibinfo{year}{1991}).

\bibitem[{\citenamefont{Suhonen et~al.}(1991)\citenamefont{Suhonen, Khadkikar,
  and Faessler}}]{Suh91}
\bibinfo{author}{\bibfnamefont{J.}~\bibnamefont{Suhonen}},
  \bibinfo{author}{\bibfnamefont{S.~B.} \bibnamefont{Khadkikar}},
  \bibnamefont{and} \bibinfo{author}{\bibfnamefont{A.}~\bibnamefont{Faessler}},
  \bibinfo{journal}{Nucl. Phys. A} \textbf{\bibinfo{volume}{529}},
  \bibinfo{pages}{727} (\bibinfo{year}{1991}).

\bibitem[{\citenamefont{Pantis and Vergados}(1990)}]{Pan90}
\bibinfo{author}{\bibfnamefont{G.}~\bibnamefont{Pantis}} \bibnamefont{and}
  \bibinfo{author}{\bibfnamefont{J.~D.} \bibnamefont{Vergados}},
  \bibinfo{journal}{Phys. Lett. B} \textbf{\bibinfo{volume}{242}},
  \bibinfo{pages}{1} (\bibinfo{year}{1990}).

\bibitem[{\citenamefont{Kami\'{n}ski and Faessler}(1991)}]{Kam91}
\bibinfo{author}{\bibfnamefont{W.~A.} \bibnamefont{Kami\'{n}ski}}
  \bibnamefont{and} \bibinfo{author}{\bibfnamefont{A.}~\bibnamefont{Faessler}},
  \bibinfo{journal}{Nucl. Phys. A} \textbf{\bibinfo{volume}{529}},
  \bibinfo{pages}{605} (\bibinfo{year}{1991}).

\bibitem[{\citenamefont{Civitarese et~al.}(1987)\citenamefont{Civitarese,
  Faessler, and Tomoda}}]{Civ86}
\bibinfo{author}{\bibfnamefont{O.}~\bibnamefont{Civitarese}},
  \bibinfo{author}{\bibfnamefont{A.}~\bibnamefont{Faessler}}, \bibnamefont{and}
  \bibinfo{author}{\bibfnamefont{T.}~\bibnamefont{Tomoda}},
  \bibinfo{journal}{Phys. Lett. B} \textbf{\bibinfo{volume}{194}},
  \bibinfo{pages}{11} (\bibinfo{year}{1987}).

\bibitem[{\citenamefont{Grotz and Klapdor}(1985)}]{Gro85}
\bibinfo{author}{\bibfnamefont{K.}~\bibnamefont{Grotz}} \bibnamefont{and}
  \bibinfo{author}{\bibfnamefont{H.~V.} \bibnamefont{Klapdor}},
  \bibinfo{journal}{Phys. Lett. B} \textbf{\bibinfo{volume}{157}},
  \bibinfo{pages}{242} (\bibinfo{year}{1985}).

\bibitem[{\citenamefont{Balian and Brezin}(1969)}]{Bal69}
\bibinfo{author}{\bibfnamefont{R.}~\bibnamefont{Balian}} \bibnamefont{and}
  \bibinfo{author}{\bibfnamefont{E.}~\bibnamefont{Brezin}},
  \bibinfo{journal}{Nuovo Cimento} \textbf{\bibinfo{volume}{64}},
  \bibinfo{pages}{37} (\bibinfo{year}{1969}).

\bibitem[{\citenamefont{\v{S}imkovic et~al.}(1999)\citenamefont{\v{S}imkovic,
  Pantis, Vergados, and Faessler}}]{Sim99}
\bibinfo{author}{\bibfnamefont{F.}~\bibnamefont{\v{S}imkovic}},
  \bibinfo{author}{\bibfnamefont{G.}~\bibnamefont{Pantis}},
  \bibinfo{author}{\bibfnamefont{J.~D.} \bibnamefont{Vergados}},
  \bibnamefont{and} \bibinfo{author}{\bibfnamefont{A.}~\bibnamefont{Faessler}},
  \bibinfo{journal}{Phys. Rev. C} \textbf{\bibinfo{volume}{60}},
  \bibinfo{pages}{055502} (\bibinfo{year}{1999}).

\bibitem[{\citenamefont{Terasaki and Engel}(2010)}]{Ter10}
\bibinfo{author}{\bibfnamefont{J.}~\bibnamefont{Terasaki}} \bibnamefont{and}
  \bibinfo{author}{\bibfnamefont{J.}~\bibnamefont{Engel}},
  \bibinfo{journal}{Phys. Rev. C} \textbf{\bibinfo{volume}{82}},
  \bibinfo{pages}{034326} (\bibinfo{year}{2010}).

\bibitem[{\citenamefont{Dobaczewski et~al.}(1984)\citenamefont{Dobaczewski,
  Flocard, and Treiner}}]{Dob84}
\bibinfo{author}{\bibfnamefont{J.}~\bibnamefont{Dobaczewski}},
  \bibinfo{author}{\bibfnamefont{H.}~\bibnamefont{Flocard}}, \bibnamefont{and}
  \bibinfo{author}{\bibfnamefont{J.}~\bibnamefont{Treiner}},
  \bibinfo{journal}{Nucl. Phys. A} \textbf{\bibinfo{volume}{422}},
  \bibinfo{pages}{103} (\bibinfo{year}{1984}).

\bibitem[{\citenamefont{Blazkiewicz et~al.}(2005)\citenamefont{Blazkiewicz,
  Oberacker, Umar, and Stoitsov}}]{Bla05}
\bibinfo{author}{\bibfnamefont{A.}~\bibnamefont{Blazkiewicz}},
  \bibinfo{author}{\bibfnamefont{V.~E.} \bibnamefont{Oberacker}},
  \bibinfo{author}{\bibfnamefont{A.~S.} \bibnamefont{Umar}}, \bibnamefont{and}
  \bibinfo{author}{\bibfnamefont{M.}~\bibnamefont{Stoitsov}},
  \bibinfo{journal}{Phys. Rev. C} \textbf{\bibinfo{volume}{71}},
  \bibinfo{pages}{054321} (\bibinfo{year}{2005}).

\bibitem[{\citenamefont{Terasaki et~al.}(2005)\citenamefont{Terasaki, Engel,
  Bender, Dobaczewski, Nazarewicz, and Stoitsov}}]{Ter05}
\bibinfo{author}{\bibfnamefont{J.}~\bibnamefont{Terasaki}},
  \bibinfo{author}{\bibfnamefont{J.}~\bibnamefont{Engel}},
  \bibinfo{author}{\bibfnamefont{M.}~\bibnamefont{Bender}},
  \bibinfo{author}{\bibfnamefont{J.}~\bibnamefont{Dobaczewski}},
  \bibinfo{author}{\bibfnamefont{W.}~\bibnamefont{Nazarewicz}},
  \bibnamefont{and} \bibinfo{author}{\bibfnamefont{M.}~\bibnamefont{Stoitsov}},
  \bibinfo{journal}{Phys. Rev. C} \textbf{\bibinfo{volume}{71}},
  \bibinfo{pages}{034310} (\bibinfo{year}{2005}).

\bibitem[{\citenamefont{Bartel et~al.}(1982)\citenamefont{Bartel, Quentin,
  Brack, Guet, and H{\aa}kansson}}]{Bar82}
\bibinfo{author}{\bibfnamefont{J.}~\bibnamefont{Bartel}},
  \bibinfo{author}{\bibfnamefont{P.}~\bibnamefont{Quentin}},
  \bibinfo{author}{\bibfnamefont{M.}~\bibnamefont{Brack}},
  \bibinfo{author}{\bibfnamefont{C.}~\bibnamefont{Guet}}, \bibnamefont{and}
  \bibinfo{author}{\bibfnamefont{H.-B.} \bibnamefont{H{\aa}kansson}},
  \bibinfo{journal}{Nucl. Phys. A} \textbf{\bibinfo{volume}{386}},
  \bibinfo{pages}{79} (\bibinfo{year}{1982}).

\bibitem[{\citenamefont{Dobaczewski et~al.}(1996)\citenamefont{Dobaczewski,
  Nazarewicz, Werner, Berger, Chinn, and Decharg\'{e}}}]{Dob96}
\bibinfo{author}{\bibfnamefont{J.}~\bibnamefont{Dobaczewski}},
  \bibinfo{author}{\bibfnamefont{W.}~\bibnamefont{Nazarewicz}},
  \bibinfo{author}{\bibfnamefont{T.~R.} \bibnamefont{Werner}},
  \bibinfo{author}{\bibfnamefont{J.~F.} \bibnamefont{Berger}},
  \bibinfo{author}{\bibfnamefont{C.~R.} \bibnamefont{Chinn}}, \bibnamefont{and}
  \bibinfo{author}{\bibfnamefont{J.}~\bibnamefont{Decharg\'{e}}},
  \bibinfo{journal}{Phys. Rev. C} \textbf{\bibinfo{volume}{53}},
  \bibinfo{pages}{2809} (\bibinfo{year}{1996}).

\end{thebibliography}

\end{document}